# Super-Planckian Radiative Heat Transfer between Macroscale Plates with Vacuum Gaps Down to 190 nm Directly Created by SU-8 Posts and Characterized by Capacitance Method

Xiaoyan Ying, Payam Sabbaghi, Nicole Sluder, and Liping Wang[*]

School for Engineering of Matter, Transport & Energy
Arizona State University, Tempe, AZ 85287, USA

[*]Corresponding Author: liping.wang@asu.edu

**ABSTRACT**

In this work we experimentally demonstrated the near-field thermal radiation enhancement over the blackbody limit by 11 times between highly doped silicon chips with 1×1 cm$^2$ size at a vacuum gap distance of 190±20 nm under a temperature difference of 74.7 K above room temperature. SU-8 polymer posts, which significantly reduced the conduction less than 6% of the total heat transfer due to its low thermal conductivity, were carefully fabricated with different heights to directly create vacuum gaps from 507±47 nm down to 190±20 nm precisely determined in-situ by capacitance measurement. Experimental results were validated by theoretical calculations based on fluctuational electrodynamics, which revealed the enhancement mechanism mainly as coupled surface plasmon polariton. The experimental method developed here will facilitate the potential applications of near-field radiative devices made of electrically conductive materials like metals, graphene, and transparent conductive oxide besides heavily doped semiconductors for thermal energy conversion, radiative thermal rectification, and radiative heat modulation.

Keywords: near-field thermal radiation, surface plasmon polariton, doped silicon.

Thermal agitation of charges inside a body at finite temperature above zero degree can emit propagating waves related to far-field thermal radiation as well as evanescent waves associated with near-field thermal radiation.[1,2] While Planck's law well describes the blackbody radiation, it is restricted to the cases where distances between bodies are greater than the characteristic thermal wavelength.[1,2] In the near-field regime however, the radiative heat flux can exceed the blackbody limit by a few orders of magnitude due to coupling of evanescent waves or so-called photon tunneling within the subwavelength vacuum gap between the emitter and receiver.[3-6] Applications of near-field thermal radiation (NFTR) are numerous ranging from thermal imaging,[7-9] thermopower conversion,[10-14] radiative cooling,[15-18] to contactless heat flow management such as thermal rectification[19-22] and modulation[23-27].

Most of the recent parallel-plate NFTR experiments demonstrated enhanced heat transfer beyond blackbody limit across submicron vacuum gaps with polar materials like $SiO_2$ and $Al_2O_3$ by taking advantage of their phonon polaritons in the infrared.[28-33] Noble metals with plasma frequencies in ultraviolet and visible ranges usually do not significantly enhance NFTR around room temperature due to non-resonant near-field coupling.[34-37] On the other hand, doped semiconductors like silicon, which are widely used in microelectronics, were predicted to enhance near-field radiative heat transfer across nanometer vacuum gaps with coupled surface plasmon polaritons (SPP) in the infrared.[38-40]

One of major challenges in NFTR measurement is the creation and accurate characterization of nanometric vacuum gaps between two parallel surfaces at macroscopic scale. Watjen et al.[41] created vacuum gaps by patterned $SiO_2$ posts, whose conduction contributes about 20% ~ 50% of the total heat transfer, and experimentally found that the NFTR between two 10×10 $mm^2$ Si plates is 11 times over the blackbody limit at 200 nm gap spacing. The doping level was limited to $2\times10^{18}$ cm$^{-3}$ as the vacuum gap distance was determined ex-situ by interference spectra from spectroscopic reflection characterization. Lim et al.[42] measured the NFTR between Si strips of doping concentration $8.33\times10^{19}$ cm$^{-3}$ at a separation gap distance of ~400 nm determined by capacitance measurement from a MEMS-based platform with microfabricated electrodes. By using a bimaterial cantilever as a thermal sensor, Shi et al.[43] demonstrated tunable NFTR of doped Si with different carrier concentrations ranging from $3\times10^{14}$ cm$^{-3}$ to $2\times10^{20}$ cm$^{-3}$ at separation gap distances down to 60 nm from a glass microsphere. Previous NFTR experiments with macroscopic surface areas that determined the gap distance with optical interference method were limited to semitransparent samples like SiO$_2$[30,33] or lightly-doped Si,[41] while fabrication of the MEMS structures and electrodes increases the complexity during sample preparation.[42]

In this paper we report a near-field thermal radiation measurement between optically-opaque electrically-conductive samples at centimeter sizes separated by vacuum gap distances down to 190 nm directly created by SU-8 polymer posts and characterized in-situ by capacitance measurement. *P*-type boron doped silicon chips at a size of $1\times1$ cm$^2$ with thickness of 525±20 μm were used in this study, while the resistivity was measured by a four-probe station (Signatone Pro-4) to be 0.0051 Ω-cm, suggesting a doping level of $2\times10^{19}$ cm$^{-3}$.[39] **Figure 1(a)** presents the theoretical near-field radiative heat flux between Si plates at various doping levels from $2\times10^{17}$ cm$^{-3}$ to $2\times10^{21}$ cm$^{-3}$ under emitter and receiver temperatures of 400K and 300 K based on fluctuational electrodynamics (see Methods for more details), where a doping level of $2\times10^{19}$ cm$^{-3}$ was predicted to achieve the highest heat flux at vacuum gaps smaller than 200 nm. However, the penetration depth for Si with doping level of $2\times10^{19}$ cm$^{-3}$ is only about 20 μm at angular frequency of $5\times10^{14}$ rad/s (i.e., 3.77 μm in wavelength) as shown in **Figure 1(b)**. Therefore, spectroscopic method could not be used for measuring the vacuum gaps as light could not penetrate the heavily doped Si wafer of a few hundred micrometers thick.

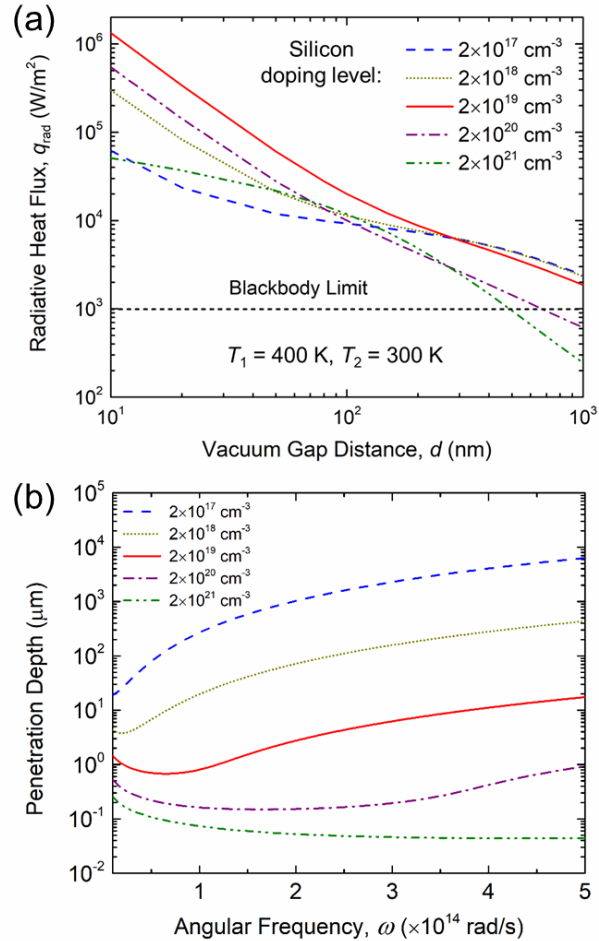

**Figure 1. Predicted NFTR between doped Si plates and silicon penetration depth at different doping levels.** (a) Near-field radiative heat flux between silicon plates at different doping levels at vacuum gaps from 10 nm to 1 μm calculated by fluctuational electrodynamics. The emitter and receiver temperatures are considered as 400 K and 300 K, respectively. The blackbody limit is also presented for comparison. (b) Penetration depth in silicon at different doping levels as a function of angular frequency calculated based on the dielectric function model of doped Si in Ref. 39.

**RESULTS AND DISCUSSION**

We measured the parallel-plate capacitance between two heavily-doped silicon samples and precisely determined the vacuum gap distance in-situ during near-field thermal measurements. As depicted in **Figure 2(a)**, samples are intentionally misaligned such that the sample corners are exposed for attaching electrical wires for the capacitance measurement. Before stacking the doped Si emitter onto the receiver sample, a total number of $N = 52$ evenly distributed SU-8 posts with diameter of $D = 3$ μm were fabricated on the receiver Si chip for creating the submicron vacuum gaps after a series of careful sample cleaning and photolithography processes (see Methods for details). SU-8 photoresist was chosen for its excellent mechanical strength as well as low thermal conductivity (i.e., $k_{SU8} = 0.2$ W·m$^{-1}$·K$^{-1}$) to reduce the conduction heat transfer via the posts. Different vacuum gap distances were achieved by certain height of SU-8 posts ranging from 500 nm down to 200 nm by well-controlled plasma etching process. **Figure 2(b)** shows a picture of the emitter sample being mounted onto the receiver with SU-8 posts. A pressure of 316.8 Pa was applied onto the emitter by the total weight (i.e., 3.23 g) of a copper heat spreader and a ceramic heater with the same size of 1×1 cm$^2$ loaded on top, which also hold the emitter in place and help compensate the wafer bow. Thin bonding wires were carefully attached with silver paste onto the corner of the receiver sample surface and onto the top copper plate, which was electrically connected to the emitter sample via carbon tape for the gap capacitance measurement with a digital multimeter (Tektronix DMM4050). The overlapping area $A_{gap}$ between the emitter and receiver samples was estimated from the optical image to be approximately 70% of the sample size, while 10% error for the area was taken into account during the error propagation analysis.

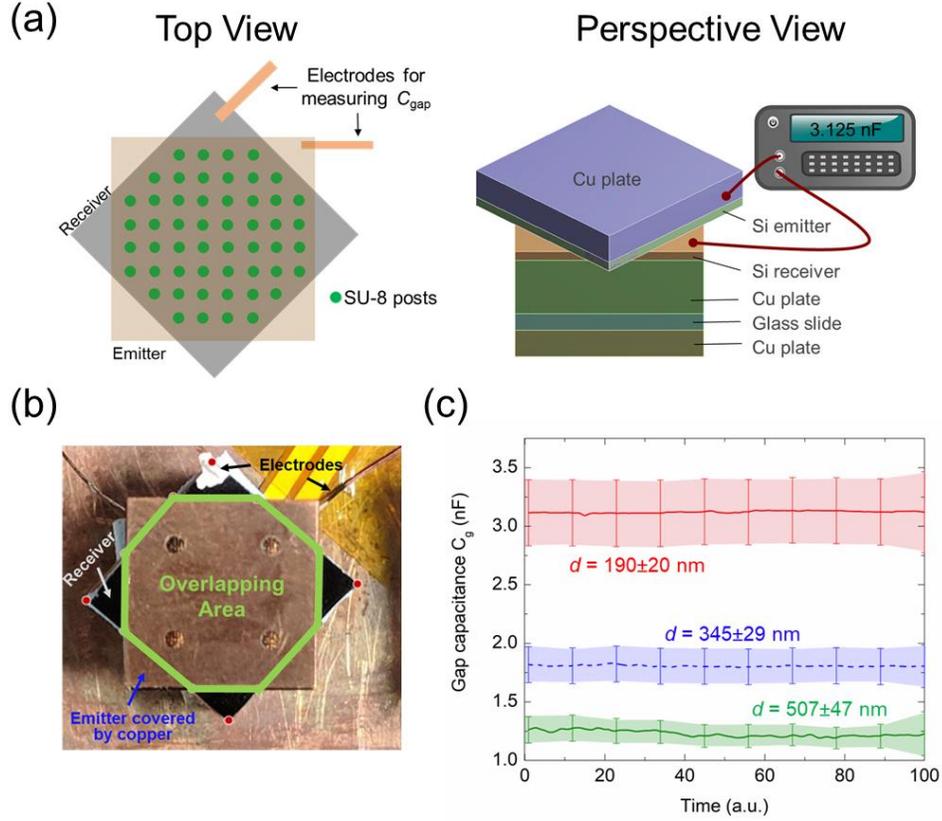

**Figure 2. Capacitance measurement for sub-micron gap distance determination.** (a) Schematics to illustrate how the vacuum gaps are created by fabricated SU-8 posts (top view), and how the gap capacitance is measured with electrodes (perspective view). (b) Photo of misaligned emitter and receiver samples after stacking with electrodes for gap capacitance measurement, where overlapping area is delineated by green lines. (c) Measured gap capacitance at three different vacuum distances with error bar indicating the standard deviation from three emitter-receiver pairs used for near-field measurements.

The vacuum gap distance can be determined as $d = \frac{\varepsilon_0 A_{\text{gap}}}{C_{gap}}$, where $C_{\text{gap}}$ is the measured gap capacitance and $A_{\text{gap}}$ is the overlapping area. **Figure 2(c)** presents the measured gap capacitance values of mounted emitter and receiver samples separated by SU-8 posts with three different heights for the NFTR experiments, where the corresponding gap distances were found to be $d$ = 507±47 nm, 345±29 nm, and 190±20 nm. The gap distance error is considered as one standard deviation of three independent measurements from three emitter-receiver pairs prepared in the same fabrication batch. Note that the gap distances obtained from capacitance measurements agree well with the heights of fabricated SU-8 posts measured by a profilometer (i.e., 508 nm, 356 nm, and 205 nm respectively). While the relative difference is less than 10%, it is more reasonable to use the gap distance directly measured from capacitance across the actual vacuum gap where the NFTR occurs between mounted emitter and receiver samples, rather than to simply take the SU-8 heights from the profilometer measurement to be the gap distance in case of accidental wafer deformation and dust particles during sample mounting.

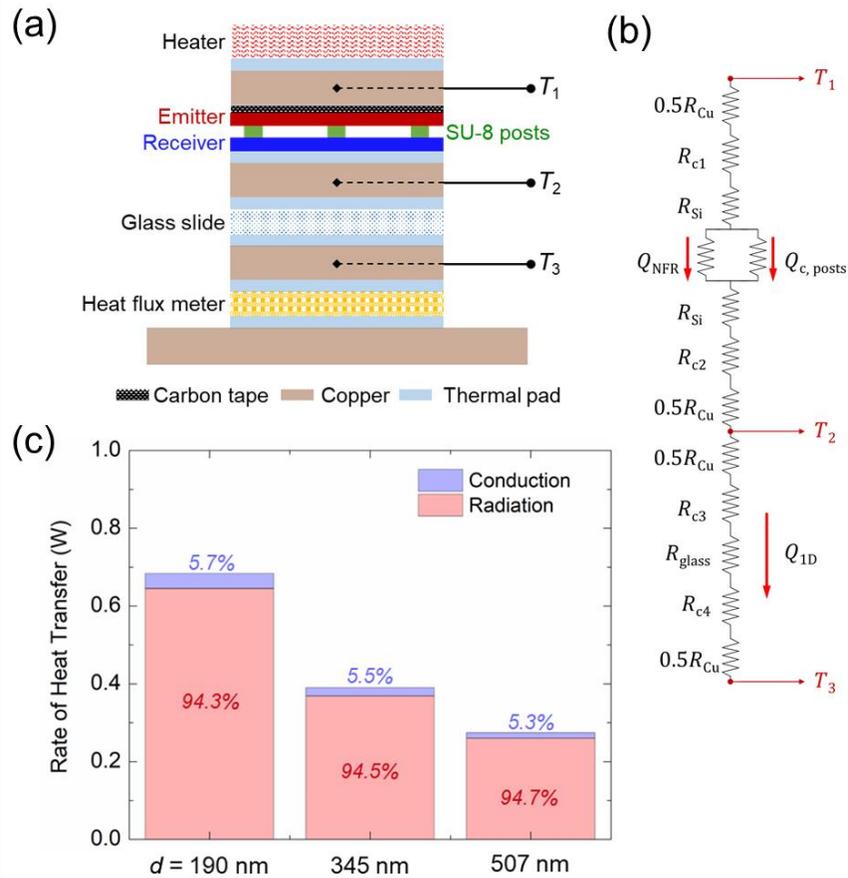

**Figure 3. NFTR measurement setup and heat transfer model.** (a) Illustration of the near-field thermal experiment setup. (b) Equivalent thermal resistance network. (c) Calculated heat transfer rate and percentages via conduction through SU-8 posts and via near-field radiation between Si plates with doping level of $2\times10^{19}$ cm$^{-3}$ at vacuum gaps of 190 nm, 345 nm, and 507 nm.

As depicted in **Figure 3(a)**, the near-field experimental setup employs three thermistors with 0.1°C resolution to measure the emitter temperature $T_1$, receiver temperature $T_2$ above the glass slide, and temperature $T_3$ below the glass slide. Note that the temperature difference between the sample surface to the center of copper (where thermistor is located) is comparable to the thermistor resolution with estimated contact resistance $R_{c,1}$ and $R_{c,2}$ as small as 0.32 K/W, and thus is reasonably neglected here. According to the equivalent thermal resistance network shown in **Figure 3(b)**, under the steady state condition and by neglecting the convection and radiation losses from the sides, the amount of 1D heat transfer across the glass slide can be found as

$$Q_{1D} = \frac{T_2 - T_3}{R_{Cu} + R_{glass} + R_{c,3} + R_{c,4}} \tag{1}$$

where $R_{Cu} = 0.08$ K/W and $R_{glass} = 9.5$ K/W were respectively calculated conduction thermal resistances of the copper plate and glass slide based on the size, thickness and thermal conductivity, while 5% of error is taken from error propagation analysis. Thermal pads were used at all interfaces, while the contact thermal resistances at both surfaces of the glass slide (i.e., $R_{c,3}$ and $R_{c,4}$) were calibrated by a thin-film heat flux sensor (GreenTEG, gSKIN® XP-26-9C) to be around 0.32 K/W. Therefore, the experimental near-field radiative heat flux can be determined by:

$$q_{NFR,exp} = \frac{Q_{NFR,exp}}{A_{gap}} = \frac{Q_{1D} - Q_{c,posts}}{A_{gap}} \tag{2}$$

where $Q_{c,posts} = \frac{T_1 - T_2}{R_{posts}}$ is the conductive heat transfer through all the SU-8 posts with thermal resistance $R_{posts} = \frac{4d}{k_{SU8} N \pi D^2}$. Considering the case where $T_1 = 123$ °C and $T_2 = 23$ °C with 70% overlapping area between 1×1 cm² emitter and receiver samples, **Figure 3(c)** shows the calculation results of conduction $Q_{c,posts}$ and near-field radiation heat transfer $Q_{NFR,theo}$ between Si plates with doping level of $2 \times 10^{19}$ cm$^{-3}$ (see Methods for details). The conduction via the SU-8 posts is predicted to be less than 6% of total heat transfer rate at vacuum gap $d = 507$ nm, 345 nm, and 190 nm. The significantly reduced conduction with SU-8 posts would help with more precise measurement of NFTR as the dominant heat transfer mode across the gap, because the actual amount of conduction heat transfer via SU-8 posts cannot be experimentally determined during the thermal measurement.

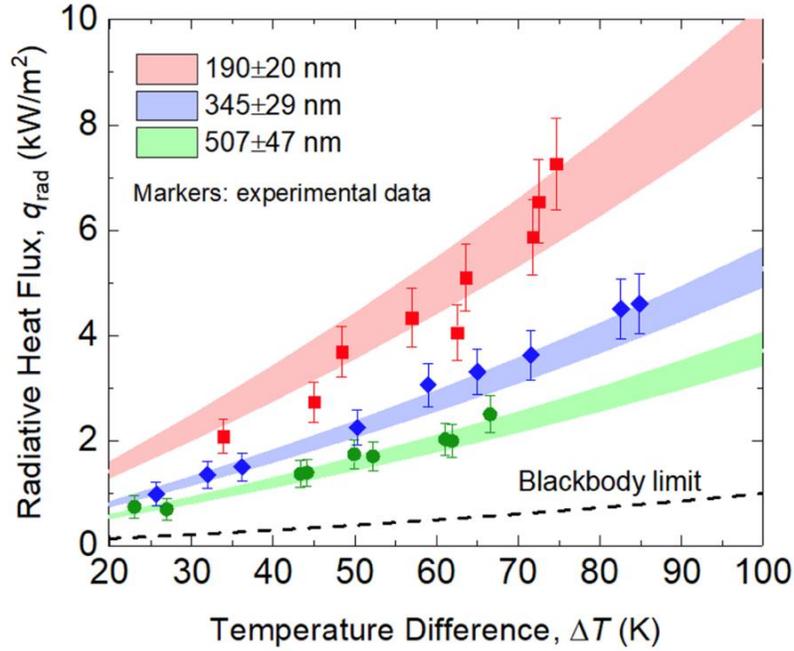

**Figure 4. Experimental and theoretical near-field radiative heat flux $q_{NFR}$ at three different vacuum gap distances as a function of the emitter-receiver temperature difference $\Delta T = T_1 - T_2$.** Filled symbols are experimental results with error bars from error propagation analysis. The colored bands represent the theoretical calculation based on fluctuational electrodynamics, while the bandwidth considers one standard deviation of gap distances from the capacitance measurement. Blackbody limit is also calculated for comparison.

At vacuum pressure lower than 0.1 Pa where convection can be neglected, multiple near-field thermal measurements were carried out at three different vacuum gaps, i.e., 507±47 nm, 345±29 nm, and 190±20 nm as characterized through the gap capacitance measurements, under several emitter-receiver temperature differences $\Delta T = T_1 - T_2$ up to 85 K. As shown in **Figure 4**, each marker presents one measurement with experimental near-field radiative heat flux deduced by the methods described earlier, while the error bar indicates the accuracy of each measurement calculated from the error propagation analysis (see Methods for details). Theoretical near-field radiative heat flux predicted by fluctuational electrodynamics was calculated as colored bands for validating the experimental data, whose bandwidth considers the uncertainty of gap distances from the capacitance measurements. A good agreement between experimental results and the theoretical prediction is clearly seen, and the near-field radiative heat transfer between heavily doped Si increases with smaller vacuum gaps or greater temperature differences. In particular, the measured near-field radiative heat flux reaches a value of 7260 W/m² at the vacuum gap $d = 190 \pm 20$ nm with a temperature difference of 74.7 K, resulting in 11 times enhancement over the blackbody limit.

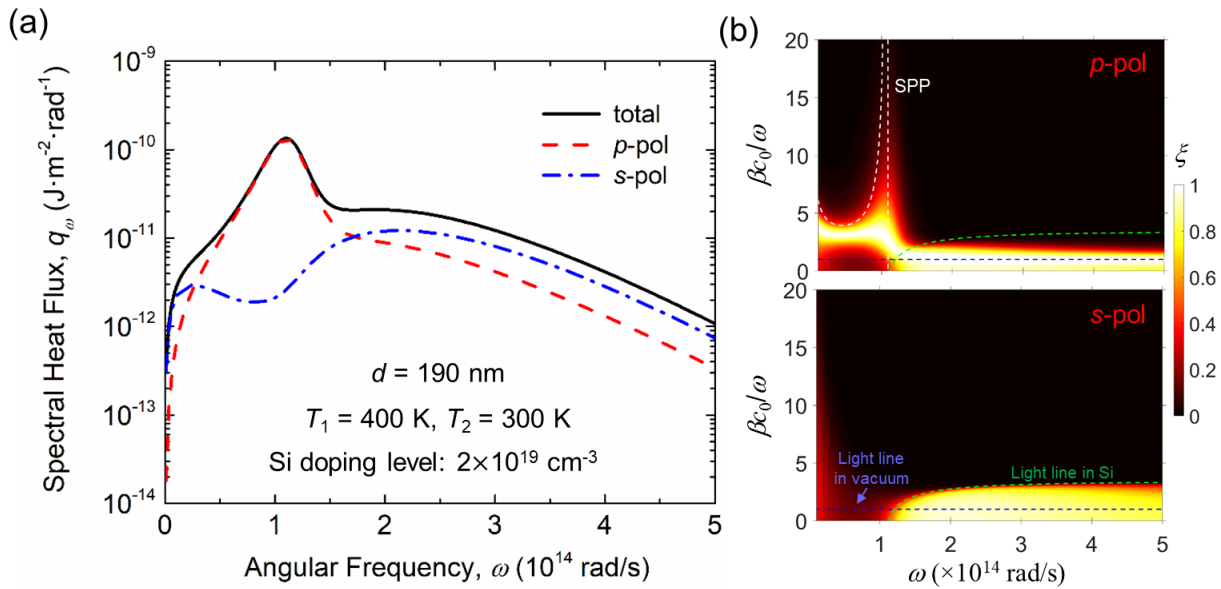

**Figure 5. Theoretical calculation of near-field radiative heat transfer between silicon plates with doping level of $2\times10^{19}$ cm$^{-3}$ at a vacuum gap $d$ =190 nm.** (a) Spectral radiative heat flux $q_\omega$, and (b) contour plots of energy transmission coefficient $\xi(\omega, \beta)$ for $s$ and $p$ polarized waves where analytical SPP dispersion curve (white dash), light line in silicon (green dash), and light line in vacuum (blue dash) are also plotted. The emitter and receiver temperatures are $T_1 = 400$ K and $T_2 = 300$ K, respectively.

In order to understand the near-field enhancement mechanism, spectral radiative heat flux was calculated at vacuum gap $d = 190$ nm with emitter temperature $T_1 = 400$ K and receiver temperature $T_2 = 300$ K for silicon samples of doping level of $2\times10^{19}$ cm$^{-1}$. As shown in **Figure 5(a)**, the total spectral heat flux peaks around angular frequency $1.1\times10^{14}$ rad/s, which is caused by the spectral enhancement from *p*-polarized waves. Note that strong SPP coupling across nanometric vacuum gaps could occur for heavily doped silicon in the infrared to result in enhanced near-field radiative transfer.[40] Moreover, the plasma frequency of silicon with $2\times10^{19}$ cm$^{-3}$ doping level is $\omega_p = 4.15\times10^{14}$ rad/s, and the asymptotic SPP frequency is predicted to be $1.15\times10^{14}$ rad/s according to $\omega_{SPP} = \frac{\omega_p}{\sqrt{\varepsilon_\infty+1}}$ with high-frequency dielectric constant $\varepsilon_\infty = 11.7$ of doped Si (see Methods for details), where the spectral enhancement occurs for *p*-polarized waves but not *s*-polarized ones.[44] This is further confirmed by the contour plots of energy transmission coefficient $\xi$ as a function of angular frequency $\omega$ and normalized parallel wavevector $\beta c_0/\omega$ (representing heat transfer channels) calculated from fluctuational electrodynamics for *p*- and *s*-polarized waves in **Figure 5(b)** (see Methods for details). Large energy transmission coefficients up to unity shown by bright contour is observed around $\omega_{SPP} = 1.15\times10^{14}$ rad/s at $\beta c_0/\omega$ values up to 20 only for *p*-polarized waves. This enhancement band is well accompanied by the analytical SPP dispersion relation curves shown by white dash lines which verify its origin. On the other hand, the *s*-polarized waves contribute more than *p*-polarized ones to the total spectral heat flux at frequencies beyond $1.6\times10^{14}$ rad/s. This can be understood from the energy transmission coefficient contour plot for *s* polarization, where strong near-field enhancement is clearly observed under the light line in silicon (i.e., $\beta c_0/\omega = \sqrt{\varepsilon'_{Si}}$ with real part of dielectric function $\varepsilon'_{Si} > 0$ at $\omega > \omega_p/\sqrt{\varepsilon_\infty}$) due to frustrated modes within silicon (i.e., propagating inside the silicon but evanescent in the vacuum).[45]

## CONCLUSION

In summary, we have demonstrated 11 times radiation heat transfer enhancement beyond the blackbody limit with near-field effect between cm-scale heavily-doped silicon at vacuum gap of 190 nm and emitter-receiver temperature difference of 74.7 K. The fabricated SU-8 posts provided a direct way to create nanometric vacuum gaps with significantly reduced conduction heat transfer. Note that SU-8 can be readily used for bonding two samples together to realize near-field radiative thermal devices. The capacitance method presented here enables precise determination of vacuum gap distances in-situ during near-field thermal measurement in particular for optically-opaque but electrically-conductive samples. With ultraflat samples and better-controlled SU-8 etching processes, it is possible to reduce the vacuum gaps down to 100 nm or less where near-field thermal radiation could be further enhanced, and thus spur the novel applications of near-field radiative thermal devices.

## METHODS

### Fabrication of Low-density SU-8 Posts

After $O_2$ plasma cleaning, silicon chips at size of 1×1 cm$^2$ were spin coated with SU-8 photoresist from Microchem with ramped speeds from 500 rpm for 5 seconds to 3000 rpm for 30 seconds ending with 12000 rpm for 2 seconds to eliminate edge effect. The samples were then placed on a hotplate for soft bake at 95℃ for one minute, followed by UV exposure at 80 mJ/cm$^2$ by an OAI 808 aligner through a well-designed photomask to obtain low-density SU-8 posts. Post-exposure baking at 95℃ for one minute was performed and then SU-8 was developed for 40 sec. Hard baking at 150℃ for five minutes was conducted to ensure good mechanical stability. SU-8 post height was measured by a profilometer. Desired height of SU-8 posts was obtained by reactive ion etching of the photoresist at 100 nm/min with well-controlled etching time and gas flow rates.

**Dielectric Function of Doped Silicon**

A Drude model was used to describe the infrared dielectric function of doped silicon as $\varepsilon(\omega) = (n + ik)^2 = \varepsilon_\infty - \frac{\omega_p^2}{\omega(\omega+i\gamma)}$, where $\varepsilon_\infty = 11.7$ is the high-frequency constant. Plasma frequency $\omega_p$ and scattering rate $\gamma$ depend on the dopant type, doping level, and temperature, which are determined by the models described in Ref. 39, while optical phonon absorption within intrinsic silicon is neglected. The penetration depth is defined as $\delta = \lambda/4\pi\kappa$, where $\lambda$ is the wavelength and $\kappa$ is the extinction coefficient calculated from the dielectric function.

**Theoretical Calculation for Near-field Radiative Heat Transfer**

Based on fluctuational electrodynamics, near-field radiative heat transfer between two radiating media with temperatures of $T_1$ and $T_2$ is expressed as[38,40]

$$q_{NFR} = \frac{1}{4\pi^2} \int_0^\infty [\Theta(\omega, T_1) - \Theta(\omega, T_2)] d\omega \int_0^\infty \xi(\omega, \beta) \beta d\beta \tag{S1}$$

where $\Theta(\omega, T) = \hbar\omega / (e^{\hbar\omega/k_B T} - 1)$ is the average energy of Planck's oscillator at temperature $T$. $\xi(\omega, \beta)$ represents the energy transmission coefficient depending on angular frequency $\omega$, and parallel-component wavevector $\beta$. The expression for $\xi(\omega, \beta)$ between two semi-infinite, homogenous, nonmagnetic materials across vacuum gap $d$ is given by[38,40]

$$\xi(\omega, \beta) = \begin{cases} \dfrac{\left(1-|r_1^s|^2\right)\left(1-|r_3^s|^2\right)}{\left|1-r_1^s r_3^s e^{i2\gamma_2 d}\right|^2} + \dfrac{\left(1-|r_1^p|^2\right)\left(1-|r_3^p|^2\right)}{\left|1-r_1^p r_3^p e^{i2\gamma_2 d}\right|^2}, & \beta < k_0 \\ \dfrac{4\,\text{Im}(r_1^s)\,\text{Im}(r_3^s) e^{-2\,\text{Im}(\gamma_2)d}}{\left|1-r_1^s r_3^s e^{i2\gamma_2 d}\right|^2} + \dfrac{4\,\text{Im}(r_1^p)\,\text{Im}(r_3^p) e^{-2\,\text{Im}(\gamma_2)d}}{\left|1-r_1^p r_3^p e^{i2\gamma_2 d}\right|^2}, & \beta > k_0 \end{cases} \tag{S2}$$

Here, the subscripts 1, 2 and 3 denote the emitter, the vacuum layer and the receiver, respectively. Transverse electric and transverse magnetic polarizations are indicated as $s$ and $p$, while $\gamma$ is the wavevector component vertical to the interface. $r$ is the Fresnel reflection coefficient at vacuum interface, and $k_0 = \omega/c_0$. The electromagnetic waves in the vacuum gap are propagating when $\beta < k_0$, or evanescent when $\beta > k_0$. The blackbody limit is calculated as $q_{BB} = \sigma(T_1^4 - T_2^4)$,[1,2] where $\sigma = 5.67 \times 10^{-8} (W/m^2 K^4)$ is the Stefan-Boltzmann constant.

**Dispersion Relation of Coupled Surface Plasmon Polaritons**

The dispersion relation for surface plasmon polariton (SPP) coupling is calculated via

$$\left|1 - r_1^p r_3^p e^{i2\gamma_2 d}\right| = 0 \tag{S3}$$

Here, the subscript 1, 2 and 3 denotes the emitter, vacuum gap and receiver mediums, $\gamma$ is the wavevector component vertical to the interface, and $r^p$ is the Fresnel reflection coefficient for $p$-polarization and $d$ represents the vacuum gap distance. In addition, collective oscillations of charges at the silicon-vacuum interface would lead to asymptote of SPP modes at large parallel wavevectors when $\varepsilon_{Si} + \varepsilon_{vacuum} = 0$.[44] By neglecting the loss or the scattering rate of doped Si for simplicity, the condition becomes $\varepsilon_\infty - \frac{\omega_p^2}{\omega^2} + 1 = 0$ with $\varepsilon_{vacuum} = 1$, and the asymptotic SPP frequency is

$$\omega_{SPP} = \frac{\omega_p}{\sqrt{\varepsilon_\infty + 1}} \tag{S4}$$

**Error Propagation Analysis**

The experimental uncertainty of the measured near-field radiative heat flux $q_{NFR}$ was evaluated based on the following error propagation analysis:

$$U = \sqrt{\left[\sum_{i=1}^n \left(\frac{\partial q_{NFR}}{\partial x_i} s_{x,i}\right)^2\right]} \tag{S5}$$

where $x_i$ is the measured variable including $R_{Cu}$, $R_{glass}$, $A_{gap}$, $T_2$ and $T_3$, while $s_{x,i}$ is the corresponding error. The partial derivatives of $q_{NFR}$ with respect to each variable is found from Eqs. 1 and 2.

**ACKNOWLEDGEMENTS**

This work was supported by Air Force Office of Scientific Research under Grant No. FA9550-17-1-0080 (X.Y. and L.W.) and by National Science Foundation (NSF) under Grant No. CBET-1454698 (P.S and L.W.). Access to the NanoFab facility at Arizona State University (ASU) for sample fabrication and characterization was supported in part by NSF contract ECCS-1542160. N.S. is grateful to the ASU Fulton Undergraduate Research Initiative program.